\documentclass[twocolumn,amsmath,amssymb,showpacs,prb]{revtex4}

\usepackage{graphicx} 	
\usepackage{bm} 		
\usepackage{amssymb}
\usepackage{amsmath}
\usepackage{epstopdf}

\begin{document}
\newcommand{\FU}{$F_U$}
\newcommand{\FJa}{$F_{J{\bf a}}$}
\newcommand{\swave}{$s$-wave}
\newcommand{\dawave}{$d_1$-wave}
\newcommand{\dbwave}{$d_2$-wave}
\newcommand{\chempot}{\ensuremath{\tilde{\mu}}}

\title{The effect of nearest neighbor spin-singlet correlations in conventional graphene SNS Josephson junctions}

\author{Annica M. Black-Schaffer}
 \affiliation{Department of Applied Physics, Stanford University, Stanford, California 94305, USA}
 \author{Sebastian Doniach}
 \affiliation{Departments of Physics and Applied Physics, Stanford University, Stanford, California 94305, USA}

\date{\today}
       
\begin{abstract}
Using the self-consistent tight-binding Bogoliubov-de Gennes formalism we have studied the effect of nearest neighbor spin-singlet bond (SB) correlations on Josephson coupling and proximity effect in graphene SNS Josephson junctions with conventional $s$-wave superconducting contacts. 
For strong enough coupling the SB correlations give rise to a superconducting state with either an extended $s$-, $d_{x^2-y^2}$-,  or $d_{xy}$-wave symmetry, or different combinations of the $d$-waves with the $d_{x^2-y^2} + id_{xy}$ state favored in the bulk. 
Despite the $s$-wave superconducting state in the contacts, the SB pairing state inside the junction has $d$-wave symmetry and clean, sharp interface junctions resemble a `bulk-meets-bulk'\ situation with very little interaction between the two different superconducting states. In fact, due to a finite-size suppression of the superconducting state, a stronger SB coupling constant than in the bulk is needed in order to achieve SB pairing in a junction. For both short clean zigzag and armchair junctions the $d$-wave state that has a zero Josephson coupling to the $s$-wave state is chosen and therefore the Josephson current {\it decreases} when a SB pairing state develops in these junctions. 
In more realistic junctions, with smoother doping profiles and atomic scale disorder at the interfaces, it is possible to achieve some coupling between the contact $s$-wave state and the SB $d$-wave states. In addition, by breaking the appropriate lattice symmetry at the interface in order to induce the other $d$-wave state, a nonzero Josephson coupling can be achieved which leads to a substantial increase in the Josephson current. 
We also report on the LDOS  of the junctions and on a lack of zero energy states at interfaces despite the unconventional order parameters, which we attribute to the near degeneracy of the two $d$-wave solutions and their mixing at a general interface. 
\end{abstract}


\maketitle
%
%
\section{Introduction}
 Since the experimental realization of a single layer of graphite, called graphene, a few years ago\cite{Novoselov04} the unusual two-dimensional Dirac-type spectrum of graphene has spurred a lot of attention and has been found to give rise to effects not usually expected in condensed matter systems.\cite{Geim07,CastroNeto07} 
However, little attention has been given to the fact that $p\pi$-bonded planar organic molecules, of which graphene is an infinite extension, show a preference for nearest neighbor spin-singlet bonds, or singlet bond (SB) in short, over polar configurations, 
as originally captured by  Pauling's\cite{Paulingbook} idea of resonating valence bonds. In a microscopic Hamiltonian this spin-singlet enhancement takes the form of an intrinsic $J {\bf S_i\cdot S_j}$ term between nearest neighbors. Baskaran\cite{Baskaran02} combined this term with band theory into a phenomenological Hamiltonian a few years ago and showed that strong enough coupling $J$ will give rise to mean-field superconductivity. The current authors\cite{Black-Schaffer07} extended this work by pointing out that the most favorable symmetry for this coupling, and with very high mean-field $T_c$, is not, as previously assumed, \swave\ but instead a two-dimensional $d$-wave which in the bulk will break time-reversal symmetry. This $d$-wave state has further been studied within the Dirac Bogoliubov-de Gennes (BdG) formalism to calculate properties of junctions between $d$-wave superconducting and normal graphene.\cite{Jiang08}
Recently many-body approaches have confirmed the possibility of realizing this $d$-wave state in doped graphene. A functional renormalization study has found that, for the physical parameters expected in graphene, the honeycomb lattice appears to flow toward a $d+id$ superconducting state\cite{Honerkamp08} and variational quantum Monte Carlo calculations on the Hubbard model has confirmed the stability of this state toward quantum phase fluctuations caused by on-site repulsion\cite{Pathak08}. In addition, by using hybrid exchange density functional theory, defect arrays in graphene have been shown to cause a ferromagnetic ground state even at room temperature for unexpectedly large defect separations.\cite{Pisani08} This state is a consequence of long-range spin polarization with alternating spin direction on adjacent atoms and was interpreted as driven by the intrinsic  spin-singlet $\pi$-band instability.

The mean-field results show that there is a quantum critical point at $J_c = 1.9t$ at half-filling, and with an estimated $J \sim t$ in graphene, no effects of SB correlations are expected in undoped graphene. But, with finite doping any $J$ will give a superconducting state, and for enough doping it is expected that $T_c$ will be measurable. However, the doping levels necessary in order to give $T_c \sim 10$~K are beyond the current technology of doping graphene by gating. It might be possible to achieve a higher degree of doping by chemical absorption on graphene, but care has to be taken not to alter any other properties of the graphene but the Fermi level. Atomic sulfur absorption on graphene seems to fulfill these requirements\cite{Black-Schaffer07} and this offers an explanation to the high-temperature superconductivity found in graphite-sulfur composites.\cite{Kopelevich06inbook, daSilva01, Moehlecke04} However, as a composite this material is not very easily modeled and 
it is therefore logical to seek simpler systems where the SB correlations might naturally be enhanced. One obvious candidate would be superconducting graphene. 
While a few other proposals exist for superconductivity in graphene\cite{Kuroki01,Onari03,Uchoa07} it has only so far been realized inside Josephson junctions. Graphene SNS Josephson junctions were recently manufactured\cite{Heersche07,Shailos07,Du08} by depositing conventional superconducting metal contacts on top of a graphene layer. A finite supercurrent, seemingly in agreement with theory,\cite{Titov06} was measured for these junctions even at the Dirac point.
Our quest here is to investigate what possible effects intrinsic SB correlations would have in a graphene SNS Josephson junction. More specifically, we are interested in the following two questions:
(1) Is it possible that the superconducting state induced from the contacts will cooperate favorably with the intrinsic SB correlations to make them stronger and therefore easier to detect, and 
(2) if the SB correlations are strong enough to cause superconductivity; what symmetries will be chosen and what are the measurable effects, such as changes in the Josephson current and interfacial zero energy states (ZESs)?

\begin{figure}[htb]
\includegraphics[scale = 1]{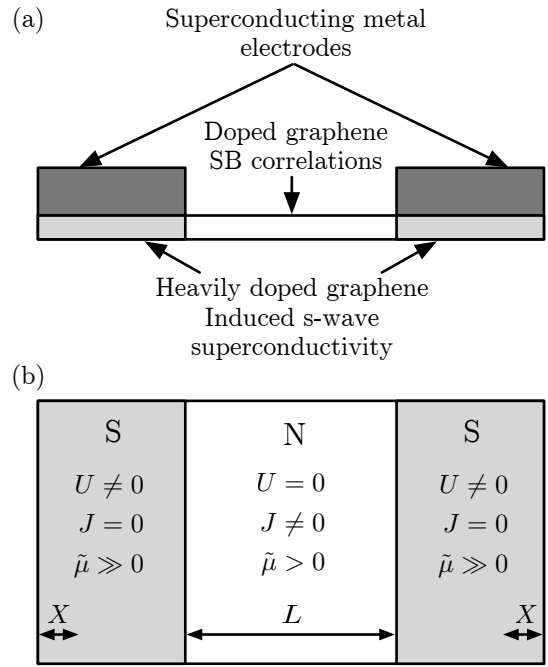}
\caption{\label{fig:junction} Schematic of (a) an experimental SNS graphene Josephson junction and (b) the model setup with input parameters: on-site pairing potential $U$, the SB coupling constant $J$, effective chemical potential $\tilde{\mu}$, length of normal region $L$, and region where the phase of the order parameter will be kept fixed $X$.}
\end{figure}

We will use a self-consistent tight-binding Bogoliubov-de Gennes (TB BdG) formulation to address these questions. The same framework was recently used by the authors\cite{Black-Schaffer08} to study graphene SNS junctions with the SB coupling switched off. This study confirmed and complemented earlier analytical results based on the non-self-consistent Dirac BdG formalism.\cite{Titov06} The benefit of using a self-consistent method is that the order parameters are allowed to vary spatially. This allows for an explicit calculation of the proximity-effect depletion and leakage of the pairing amplitude around the interface and results in a Josephson current properly calculated from the proximity effect. In addition, when the main interest is the possible coupling between different order parameters at interfaces as in the current work, a self-consistent method is a must.
More specifically, we assume the following model. The graphene sheet is assumed to be impurity-free and ballistic. The influence of the superconducting contacts deposited on top is simulated by two effects: an on-site attractive Hubbard pairing potential $U$ and heavy doping. This will give rise to a conventional, $k$-independent, \swave\ superconducting state in the S regions of the graphene and is the simplest way to simulate the effect of the superconducting contacts in the present model. Further motivation for this choice can be found in Ref.~\onlinecite{Black-Schaffer08}. The SB correlations are modeled by assuming a finite coupling $J$ in the graphene and a finite doping, though not nearly as high as in the S regions. 
Due to the significant increase in computational complexity we will not solve self-consistently for the chemical potential in the system as a function of doping but will assume fixed effective chemical potentials \chempot\ in the S and N regions. We study both clean and smooth SN interfaces as well as rough interfaces with atomic scale disorder. Experiments have indicated a high transparency of graphene SN interfaces\cite{Heersche07,Du08} so these two types of interfaces should in between them capture a realistic situation.
Figure \ref{fig:junction} shows the schematic of (a) the experimental and (b) model setup. Note that, even when we are using a large enough $J$ to achieve a SB superconducting state in N, we will, for simplicity, still call this region N, as in normal metal.

The paper is organized as follows. In the Sec.~II the TB-BdG formalism will briefly be explained and the choices for the physical parameters will be discussed. Next we will report our results. First, it is important to understand the effect of the two coupling parameters, $U$ and $J$, in the translational invariant bulk, and how their superconducting states can influence each other by induced pair amplitudes as well as Josephson coupling between them at an interface. We then  show the results for clean zigzag and armchair interfaces. Following these results, we expand the study to extended interfaces, including smoother doping profiles and interface roughness, and show that additional effects can occur. We then discuss local density of states (LDOS) results for the junctions and the presence of ZESs.
In the last section we summarize our results and discuss possible extensions and experimental opportunities.
%
%
\section{Method}
%
\subsection{Tight-binding BdG}
The method used here follows closely that of our earlier work \cite{Black-Schaffer08} where graphene SNS Josephson junctions were studied but the spin-singlet bond correlations in graphene switched off. In fact, the only difference to the formalism is the addition of the SB pairing term but for self-containment we choose to include a brief summary of the full formalism.

\begin{figure}[htb]
\includegraphics[scale = 1]{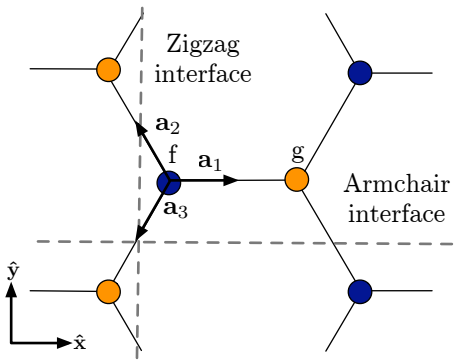}
\caption{\label{fig:graphene} (Color online) The graphene honeycomb lattice with the two different atomic sites $f$ and $g$, the three nearest neighbor vectors $\{{\bf a}_1,{\bf a}_2,{\bf a}_3\}$ (marking bonds 1, 2, and 3, respectively), and the zigzag and armchair interfaces marked.}
\end{figure}

Following the motivation in Sec.~I, we model the SNS Josephson junction using the following Hamiltonian:
%
\begin{align}
\label{eq:H_eff}
H_{\rm eff} = & -t \!\! \! \! \sum_{<{\bf i},{\bf j}>,\sigma} \!\!\! (f_{{\bf i}\sigma}^\dagger g_{{\bf j}\sigma} + g_{{\bf i}\sigma}^\dagger f_{{\bf j}\sigma}) +
\sum_{{\bf i},\sigma} \tilde{\mu}({\bf i})(f_{{\bf i}\sigma}^\dagger f_{{\bf i}\sigma} + g_{{\bf i}\sigma}^\dagger g_{{\bf i}\sigma}) \nonumber \\
& -\sum_{\bf i} U({\bf i}) (f_{{\bf i}\uparrow}^\dagger f_{{\bf i}\uparrow}f_{{\bf i}\downarrow}^\dagger f_{{\bf i}\downarrow} + g_{{\bf i}\uparrow}^\dagger g_{{\bf i}\uparrow}g_{{\bf i}\downarrow}^\dagger g_{{\bf i}\downarrow}) \nonumber \\
& -\sum_{<{\bf i},{\bf j}>}2J({\bf i})h_{\bf ij}^\dagger h_{\bf ij}
\end{align}
Here $f_{{\bf i}\sigma}^\dagger$ ($g_{{\bf i}\sigma}^\dagger$) is the creation operator on the A (B)-site in cell ${\bf i}$ of the honeycomb lattice, see Fig.~\ref{fig:graphene}. 
%
The first two terms give the tight-binding band structure for the two $\pi$-bands with hopping parameter $t = 2.5$~eV and a site-dependent effective chemical potential $\tilde{\mu}({\bf i})$ which will determine the local occupancy. $\tilde{\mu} = 0$ corresponds to the Dirac point.
The third term models the influence of the superconducting metal contacts on the underlying graphene by an on-site attractive Hubbard $U$-term. This term will only be nonzero in the S regions. Finally, the last term describes the SB correlations. Here
%
\begin{align}
\label{eq:hij}
h_{\bf ij}^\dagger = \frac{1}{\sqrt{2}}(f_{{\bf i}\uparrow}^\dagger g_{{\bf j}\downarrow}^\dagger - f_{{\bf i}\downarrow}^\dagger g_{{\bf j}\uparrow}^\dagger)
\end{align}
is the spin-singlet creation operator. We have for simplicity defined the strength of the $J$-term as given by its value on the $f$-atom in each spin-singlet. The factor of 2 comes from the two atoms per unit cell. As will be motivated below, $J$ is only set to be nonzero in the N regions.
The spin-singlet coupling can also be written as
%
\begin{align}
\label{eq:RVBterm}
-Jh_{\bf i j}^\dagger h_{\bf i j} = J \left(  {\bf S}_{\bf i} {\bf \cdot S}_{\bf j} - \frac{1}{4}n_{\bf i} n_{\bf j} \right),
\end{align}
which shows the similarity between our phenomenological model and a $t$-$J$ model, though in the latter case double occupancy is explicitly forbidden, whereas here we do not impose any such restrictions.

To proceed, we use the Hartree-Fock-Bogoliubov mean-field approximation and arrive at the following quadratic Hamiltonian:
%
\begin{align}
\label{eq:H_mf}
H_{{\rm MF}} = & -t \!\! \! \!  \sum_{<{\bf i},{\bf a}>,\sigma} \!\! \! (f_{{\bf i}\sigma}^\dagger g_{{\bf i+a},\sigma} + g_{{\bf i}\sigma}^\dagger f_{{\bf i-a},\sigma}\!) \nonumber \\
& + \! \sum_{{\bf i},\sigma} \tilde{\mu}({\bf i})(f_{{\bf i}\sigma}^\dagger f_{{\bf i}\sigma} + g_{{\bf i}\sigma}^\dagger g_{{\bf i}\sigma}\!) \nonumber \\
& +\sum_{\bf i} \Delta_{U}({\bf i}) (f_{{\bf i}\uparrow}^\dagger f_{{\bf i}\downarrow}^\dagger +  g_{{\bf i}\uparrow}^\dagger g_{{\bf i}\downarrow}^\dagger) + {\rm H.c.} \nonumber \\
& + \sum_{\bf i,a} \Delta_{J{\bf a}} (f_{{\bf i}\uparrow}^\dagger g_{{\bf i+a}\downarrow}^\dagger - f_{{\bf i}\downarrow}^\dagger g_{{\bf i+a}\uparrow}^\dagger) + {\rm H.c.},
\end{align}
where we have introduced the three nearest neighbor vectors ${\bf a}_i$, with $i = 1,2,3$, to index the nearest neighbor directions, see Fig.~\ref{fig:graphene}. The spatially dependent mean-field superconducting order parameters are defined as 
%
\begin{align}
\label{eq:orderU}
\Delta_{U}({\bf i}) & = -U({\bf i})\frac{\langle f_{{\bf i}\downarrow}  f_{{\bf i}\uparrow}\rangle + \langle g_{{\bf i}\downarrow}  g_{{\bf i}\uparrow}\rangle}{2} \nonumber \\
\Delta_{J{\bf a}}({\bf i}) & = -J({\bf i})\langle f_{{\bf i}\downarrow} g_{{\bf i+a}\uparrow} - f_{{\bf i}\uparrow} g_{{\bf i+a}\downarrow} \rangle.
\end{align}
The order parameter for the SB coupling is dependent on ${\bf a}_i$ and we will treat all three nearest neighbor bond directions independently in order to include all possible symmetries of this order parameter. We will use $\Delta_J = \sqrt{|\Delta_{J{\bf a}_1}|^2 + |\Delta_{J{\bf a}_2}|^2 + |\Delta_{J{\bf a}_3}|^2}$ as a measure of the strength of the SB order parameter.
In order to study proximity effects we also need the pairing amplitudes for each order parameter which are simply defined as $F_U({\bf i}) = -\Delta_U({\bf i})/U({\bf i})$ and $F_{J({\bf a})}({\bf i}) = -\Delta_{J({\bf a})}({\bf i})/(\sqrt{2} J({\bf i}))$.

The mean-field Hamiltonian can be rewritten in the standard BdG formalism (see Refs.~[\onlinecite{Zhu00,Covaci06,Black-Schaffer08}] for recent applications) which in a tight-binding formulation consists of an eigenvalue problem which needs to be solved for its positive eigenvalues $\nu$:
%
\begin{align}
\label{eq:eigproblem}
\sum_{\bf j} \left( \begin{array}{cc} H_0({\bf i,j}) & \Delta({\bf i,j}) \\ \Delta^\dagger({\bf i,j}) & -H_0({\bf i,j}) \end{array}\right)
\left( \begin{array}{c} 
u_{\bf j}^\nu  \\  v_{\bf j}^\nu
\end{array}\right) 
= E^\nu 
\left( \begin{array}{c}
u_{\bf i}^\nu \\ v_{\bf i}^\nu
\end{array}\right).
\end{align}
For the two-atom graphene unit cell the matrices $H_0$ and $\Delta$ will have dimensions $2\times 2$ and can be written as
%
\begin{align}
\label{eq:H0}
H_0({\bf i,j}) & =  \left( \begin{array}{cc}
\tilde{\mu}({\bf i})\delta_{{\bf ij}} & -t \sum_{\bf a} \delta_{\bf i+a,j} \\
-t \sum_{\bf a} \delta_{\bf i-a,j} & \tilde{\mu}({\bf i})\delta_{\bf ij}
\end{array}\right) \\
\label{eq:Delta}
\Delta({\bf i,j}) & =  \left(  \begin{array}{cc}  \Delta_{U}({\bf i})\delta_{\bf ij} & \sum_{\bf a} \Delta_{J{\bf a}}\delta_{\bf i+a,j} \\
\sum_{\bf a} \Delta_{J{\bf a}}\delta_{\bf i-a,j} &  \Delta_{U}({\bf i})\delta_{\bf ij}
\end{array}\right).
\end{align}
Thus $N$ unit cells leads to a $4N \times 4N$ eigenvalue problem of which $2N$ eigenvalues will be positive since $H_0 = H_0^\ast$ and $\Delta = \Delta^T$. 
Writing the eigenvectors explicitly as four-dimensional (per unit cell ${\bf i}$) using the new notation $(u,v) \rightarrow (u,y,v,z)$ we can write the self-consistency equations as
%
\begin{align}
\label{eq:selfconsU}
\Delta_{U}({\bf i}) & = \frac{U({\bf i})}{2} \sum_{\nu =1}^{2N} (u_{\bf i}^\nu v_{\bf i}^{\nu\ast} +  y_{\bf i}^\nu z_{\bf i}^{\nu\ast})\tanh \frac{\beta E^\nu}{2} \\
\label{eq:selfconsJ}
\Delta_{J{\bf a}}({\bf i}) & = J({\bf i}) \sum_{\nu =1}^{2N} (y_{\bf i+a}^\nu v_{\bf i}^{\nu \ast} + u_{\bf i}z_{\bf i+a}^{\nu \ast})\tanh \frac{\beta E^\nu}{2}.
\end{align}

Any structure with spatially varying \chempot, $U$, and $J$ can now be solved by starting with an initial guess for $\Delta_U$ and $\Delta_{J{\bf a}}$. After finding the $2N$ positive eigenvalues of Eq.~(\ref{eq:eigproblem}) we can compute new values  for $\Delta_U$ and $\Delta_{J{\bf a}}$ using Eqs. (\ref{eq:selfconsU}) and (\ref{eq:selfconsJ}) and repeat the process  until the differences in $\Delta_U$ and $\Delta_{J{\bf a}}$ between two subsequent iterations are less than the desired accuracy and thus self-consistency is reached.

From a computational point of view it is beneficial to reduce the size of the eigenvalue problem as much as possible. This can be achieved by assuming translational symmetry in the direction perpendicular to the junction and then applying Bloch's theorem. We have implemented both clean, smooth interfaces, with full translational symmetry along the junction interfaces, and simulated the effect of a rough interface by using an extended unit cell which includes a total of four graphene unit cells.

One of the main characteristics of a SNS junction is the Josephson current it can carry when a finite phase gradient is applied across it. Here, since the contacts are modeled with the attractive $U$-term, a phase difference needs to be established in $\Delta_U$ in order to produce any supercurrent. We achieve this by fixing the phase of $\Delta_U$ in the outermost regions, labeled $X$ in Fig.~\ref{fig:junction}, of the contacts and then solving self-consistently for both phase and amplitude in the rest of the structure. The current can then be calculated relatively straight-forward using the continuity equation and the Heisenberg equation for the electron density, see Ref.~[\onlinecite{Black-Schaffer08}] for further details.

The main output information from a self-consistent solution of the TB BdG formalism is (1) the pairing amplitudes $F_U$ and $F_{J{\bf a}}$ which will describe the proximity effect in the junction and (2) the Josephson current the junction carries when a finite phase gradient is applied across the junction.
%
%
\subsection{Simulation details}
With the directional dependence of the SB coupling, different interfacial directions can exhibit different behaviors. We study the two main interfaces for the graphene honeycomb lattice which are the zigzag and the armchair terminations, see Fig.~\ref{fig:graphene}. Interfaces in a random direction are called chiral interfaces and while we do not explicitly study these we will be able to infer general results from the two limiting cases of zigzag and armchair interfaces. 

The physical input parameters are the on-site pairing potential $U$ in S, the SB pairing potential $J$, the effective potential $\tilde{\mu}$ in S and N, the length $L$ of N, and temperature. 
For the superconducting contacts we choose the following setup: $U({\rm S}) = 3.4$~eV $= 1.36t$, $\chempot({\rm S}) = 1.5$~eV = 0.6$t$. This leads to $\Delta_{U}  = 0.1$~eV and a superconducting coherence length  $\xi \approx 50$~\AA\ which corresponds to 25~unit cells in the zigzag direction and 40~unit cells in the armchair direction. These values satisfy $\lambda_F(S) \ll \xi$, allow us to numerically investigate both the $L<\xi$ and $L>\xi$ cases, and coincide with the choice we used previously\cite{Black-Schaffer08} when investigating SNS junctions without the $J$-term. They are however quite large values for a realistic situation but smaller superconducting gaps lead to slower convergence rates and also the need for larger systems making calculations less feasible. We have checked our key results for smaller $U$ and found no significant difference. 

We choose to set $J = 0$ in the S regions. Physically this can be motivated by the fact that the graphene is here in very close proximity to a large metallic electron reservoir which provide ample screening of the on-site repulsion which in turn leads to significantly reduced effective coupling $J$. This setup is not akin to the situation in the intercalated graphites where we previously have argued\cite{Black-Schaffer07} for diminishing SB correlations. Also, as we will discuss below, from a practical point of view, any reasonably value of $J$ together with a large attractive $U$ will not change the physical state in any other way than to induce an even larger $\Delta_U$.
We initially studied junctions with the physically realistic value\cite{Baskaran02} $J = t = 2.5$~eV and with doping levels in N ranging from zero to no Fermi level mismatch (FLM) at the interfaces, i.e.~\chempot(N) = \chempot(S). This value of $J$ gives $T_c \approx 10$~K for $\chempot = 0.7$~eV in the bulk. However, as will be clear below, $J = t$ only induces a superconducting state in either very long, approaching bulk conditions, or in very heavily doped ($\chempot > 0.7$~eV) junctions. 
In order to keep the system sizes down and the doping level such that a significant FLM is still present at the interface, but still study the effect of order parameter symmetry selections and the physical consequences thereof, we have also preformed simulations with $J \approx 1.5t$ and $\chempot({\rm N}) = 0.7$~eV. Different doping levels in S and N are especially important in order to study effects on the Josephson current as no FLM in fact leaves the whole junction fully superconducting at the currently accessible junction lengths.\cite{Black-Schaffer08}

For most junctions we have for simplicity assumed that the doping profile changes abruptly from \chempot(S) to \chempot(N) at the interface. Physically this means insignificant doping leakage from the S regions to the N region. To examine the effect of this approximation we have also studied junctions with a linear doping profile drop over a few unit cells.
In terms of $L$, we have studied zigzag junctions with $L = 10, 30, 60$~unit cells (1 unit cell = 2.13 \AA) and  armchair junctions with the corresponding $L = 18, 52, 104$~unit cells (1 unit cell = 1.23 \AA).
The temperature was chosen to be $T = 10$~K throughout the work, which in comparison to $T_c$ in the S regions is effectively zero temperature.

The accuracy of the solution is determined by the choice of termination criterion for the self-consistency step, the number of $k$-points used in the Fourier transform preformed in the direction parallel to the junction, and the size of S and $X$ to ensure bulklike superconducting conditions in the S regions, all of which have been tested thoroughly.
%
%
\section{Results}
Before proceeding with the numerical results for SNS Josephson junctions it is important to understand what different symmetries the order parameter for SB pairing can take in the bulk and also how these can couple to the \swave\ symmetric superconducting state induced from the contacts.
%
\subsection{Order parameters in the bulk}
In the clean limit, the bulk has translational symmetry in both spatial directions. This reduces Eq.~(\ref{eq:eigproblem}) to a $1 \times 1$ matrix eigenvalue problem, which needs to be solved for each reciprocal vector. We will here  briefly review the different superconducting states possible when $U$ and/or $J \neq 0$, results that have all been published elsewhere, as well as how nonzero pairing amplitudes can be induced through order parameter interaction.
\subsubsection{$U \neq 0, J = 0$}
This case corresponds to the attractive Hubbard model on the graphene honeycomb lattice which mean-field solution is straightforward to obtain (see e.g.~Ref.~[\onlinecite{Zhao06}]). At half-filling, i.e.~$\chempot = 0$, the zero density of states leads to a quantum critical point such that $U > U_c = 2.3t$ is necessary for a superconducting state. However, at finite doping any $U$ will give a finite superconducting $T_c$. Since the pairing is on-site, no $k$-dependence is present and the order parameter has a conventional $s$-wave symmetry throughout the Brillouin zone. 
\subsubsection{$U = 0, J \neq 0$}
$J \neq 0$ represents the case of SB correlations which are present in all $p\pi$-bonded planar organic molecules. By treating the three nearest neighbor bonds independently there are three degrees of freedom for the superconducting order parameter $\Delta_{J{\bf a}}$. 
The mean-field solution of this model has been studied by the authors in a previous paper\cite{Black-Schaffer07} and it was found that the solution can either have an extended $s$-wave symmetry or belong to a two-dimensional $d$-wave symmetry class. The superconducting order parameter in graphene will belong to the $D_6$ class\cite{Jiang08} with the $s$-wave belonging to the $A_{1}$ irreducible representation and the $d$-waves to the $E_{2}$ representation. 
The extended $s$-wave is found when $\Delta_{J{\bf a}} = (\Delta_{J{\bf a}_1},\Delta_{J{\bf a}_2},\Delta_{J{\bf a}_3})\propto (1,1,1)$, i.e.~when all three bonds have the same weight. It is an extended $s$-wave in the sense that the order parameter has the same symmetry as the band structure, $|\sum_{\bf a}e^{i{\bf k}\cdot{\bf a}}|$, in the band structure Brillouin zone, i.e.~in the reciprocal space where the kinetic energy term is diagonal. This \swave\ has only nodes at the corners of the Brillouin zone.
The $d$-wave solution is two-dimensional and is spanned by $\{(2,-1,-1),(0,1,-1)\}$. Here the $(2,-1,-1)$ choice has a $d_{x^2-y^2}$ symmetry in the band structure Brillouin zone whereas the $(0,1,-1)$ has a $d_{xy}$ symmetry, see Fig.~\ref{fig:scop}. For simplicity we will use the notation $d_1 = d_{x^2-y^2}$ and $d_2 = d_{xy}$.
At $T_c$ they are degenerate, but below $T_c$ the complex combination $d_{1} + id_{2}$ has the lowest free energy, which means that the superconducting state breaks time-reversal symmetry. 
At half-filling the $s$-wave and $d$-wave solutions are degenerate and, as in the case of the attractive Hubbard model, there exists a quantum critical point at $J_c = 1.9t$ such that $J>J_c$ is necessary for a mean-field superconducting state. This is significantly higher than the estimated $J \sim t$. However, for finite doping any $J$ will give a superconducting state and the $d$-wave solutions will be favored for any reasonable values of $J$ and \chempot. A typical value of the mean-field $T_c$ for $J = t$, $\chempot = 0.7$~eV is 10~K.
\subsubsection{$U \neq 0, J \neq 0$}
When both $U$ and $J$ are nonzero the order parameters $\Delta_U$ and $\Delta_{J{\bf a}}$ have the possibility of mixing. Since we are considering $U \neq 0$ only in the S regions where also the doping level is very high, we can simplify the Hamiltonian in this case to only include one of the $\pi$-bands. It is then straightforward to show that among the four possible order parameter states the two $d$-wave states from the $J$-term are left unchanged by a finite $U$ and also $\Delta_U = 0$ for these states. The remaining two states are mixtures of the two $s$-wave states, i.e.~$\Delta_U \neq 0$ and $\Delta_{J{\bf a}} \propto (1,1,1)$.
In terms of $T_c$ the mixed $s$-waves states are heavily benefiting from a finite $U$. In fact, for $J = t$ and strong enough $U$ to create a stable induced $s$-wave from the contacts, one of the mixed $s$-wave states has a higher $T_c$ than the $d$-wave states. This leads to a situation in the S regions where, from a simulations point of view, having $J$ zero or nonzero is irrelevant, especially since, as seen in Sec.~IIIA4, a finite attractive $U$ will always induce a finite $s$-wave pairing amplitude $F_{J{\bf a}}$ so even proximity-type effects are not ignored by setting $J = 0$. For simplicity we therefore, unless otherwise stated, never assign both $U$ and $J$ nonzero at the same site. 
\subsubsection{Induced pairing amplitudes}
In the above sections we discussed the different order parameters possible when either $U$ or $J$ are nonzero. When studying the proximity effect we are, however, also interested in induced pairing amplitudes from one order parameter to the other.
For $U \neq 0$ it is straightforward to show that a finite $F_{J{\bf a}}$ with $s$-wave symmetry will always be produced. Thus inside the S regions where the superconducting contacts produce an on-site $s$-wave state, a SB $s$-wave amplitude will also be present.
Conversely, a finite SB $s$-wave state will induce a finite on-site pairing amplitude. However, a finite SB $d$-wave state has a zero overlap with the on-site $s$-wave state and, therefore, as soon as a $d$-wave state is chosen when $J \neq 0$, there is no effect on the on-site pairing amplitude in the bulk. This can easily be understood from the simple fact that the sum over the whole Brillouin zone of a function with sixfold symmetry multiplied with a function of fourfold symmetry is zero.
%
\subsection{Order parameter coupling at interfaces}
So far we have only discussed what happens in the translational invariant bulk but in order to interpret the SNS junction results it is also important to understand how two bulk order parameters would couple to each other if lined up at an interface.
The most relevant situation for the SNS structures studied here is obviously the possible coupling between an on-site, $k$-independent, $s$-wave state on one side and a SB $s$- or $d$-wave state on the other side of the interface. 

This situation corresponds to the Josephson effect in a SS$^\prime$ junction. While we do not have an explicit insulator or barrier in between the two superconductors, the FLM at the interface will act as an effective barrier, as will be clear from the results below. 
Quite generally, if we write the order parameters on each side of the junction as
%
\begin{align}
\label{eq:orderSS}
n_k = \left\{ \begin{array} {c} \tilde{n}_k^A e^{i\phi_k^A}{\rm,~for~S'} \\  \tilde{n}_k^B e^{i\phi_k^B} {\rm,~for~S,} \end{array} \right.
\end{align}
where $k = 1,2$ allows for two component order parameters, then the supercurrent density can be calculated using Ginzburg-Landau theory as\cite{Zhu98,Stefanakis01}
%
\begin{align}
\label{eq:GLcurrent}
J = \sum_{k,l = 1}^2 J_{ckl}\sin(\phi_k^B - \phi_k^A),
\end{align}
where $J_{ckl} \propto \tilde{n}_l^A \tilde{n}_k^B$. The consequences in our case depend on the dimensionality of the SB order parameter. For the extended \swave\ or a one component $d$-wave state, the magnitude of the coupling depends on the symmetry of that state at the interface and the current-phase relation has the usual $\sin(\Delta \phi)$ form, where $\Delta \phi$ is the relative phase of the two superconductors. For the time-reversal symmetry breaking $d_1+id_2$-wave the current can instead be written as $J_1 \sin(\Delta \phi) + J_2 \sin(\Delta \phi + \pi/2)$ where the terms $J_i$ depend on the symmetry of the $d_1$- and \dbwave\ components, respectively, at the interface.

\begin{figure}[htb]
\includegraphics[scale = 0.9]{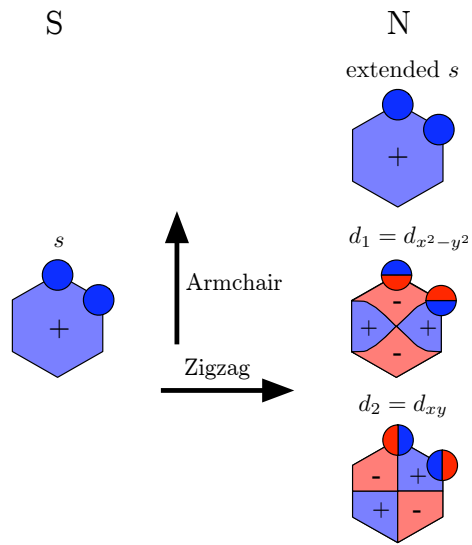}
\caption{\label{fig:scop} (Color online) Order parameter symmetry in the Brillouin zone and projected to the Fermi surface at finite doping; positive order parameter (blue), negative (red). On the S side (left) the order parameter is conventional $s$-wave but on the N side (right) for strong enough $J$ we can have either extended $s$-, $d_1$-, $d_2$-wave or any complex combination of the $d$-waves. The directions of transport for zigzag and armchair interfaces are indicated with large arrows. Note that the schematic does not include a possible additional overall phase difference between the S and N sides.}
\end{figure}

For our purpose it is enough to understand the possible couplings in a SS$^\prime$ junction on a schematic level only. Figure \ref{fig:scop} shows the three different cases; an on-site $s$-wave on the left-hand side and the three different SB order parameters on the right-hand side. As for any electronic property it is most important to consider the situation around the Fermi surface, which in doped graphene consists of two nonequivalent circles at two neighboring corners of the Brillouin zone. Figure \ref{fig:scop} explicitly shows that while the $d_1$- and the \dbwave\ states have a four-fold symmetry in the whole Brillouin zone they have effective $p_y$- and $p_x$-wave symmetries on the Fermi surface, respectively. Note that this is the same $d$-wave state as discussed in Ref.~[\onlinecite{Jiang08}] but there the low energy properties are instead given when using atomic operators. We have instead used the band structure basis where the kinetic energy is diagonal.
%
%
As can be seen in the figure, for transport through a zigzag interface, a partial overlap of the order parameters on the Fermi surface exists for the $s$- and $d_2$-states as long as we exclude intervalley scattering but the overlap is zero for the $d_{1}$-state. The opposite relation is true for the $d$-wave states in transport through an armchair interface. We would therefore expect the Josephson supercurrent through a zigzag SNS junction to be enhanced if the N region is in a $d_{2}$- or extended $s$-wave SB superconducting state or similarly for an armchair junction if the N region has $d_1$- or extended $s$-wave symmetry, but that no effect will be seen for the other choices of $d$-wave symmetry. 
For the two component $d_1+id_2$ order parameter there is finite overlap only for the $d_1$-part at an armchair interface and only for the $d_2$-part at a zigzag interface. However, note that in this particular case, due to symmetry, the extra $\pi/2$ phase shift will effectively be part of the overall phase difference. 

It might be worth noting before closing this section, that the above graphical representation based on Ginzburg-Landau theory only treats the coupling between the two superconducting order parameters to first order. It has been shown by treating the tunneling process to all orders in perturbation theory that  even if the first-order term, proportional to $\sin(\Delta \phi)$, disappears for a SS$^\prime$ junction there might still be contributions from higher order terms, starting with a $\sin(2\Delta \phi)$ term.\cite{Tanaka94}
%
%
\subsection{SNS junctions with clean interfaces}
With the preparation of the above sections, the interpretation of the numerical results for SNS junctions is quite straightforward. We start with examining clean interfaces where the doping profile drops sharply from $\chempot({\rm S})$ to $\chempot({\rm N})$ at the interface and the $U$-term is abruptly interchanged for a $J$-term. 

\begin{figure}[htb]
\includegraphics[scale = 0.65]{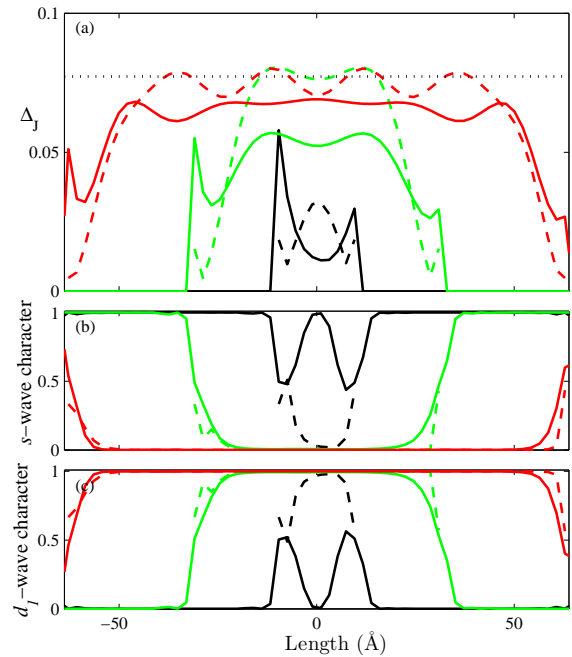}
\caption{\label{fig:azz} (Color online) (a) $\Delta_J$, (b) $s$-wave, and (c) $d_1$-wave character of $\Delta_J$ for zigzag slabs (dashed) and SNS junctions (solid) for $L = 10$ (black), 30 (green), and 60 (red) unit cells when $J = 4$~eV and $\chempot({\rm N}) = 0.7$~eV. The bulk value of $\Delta_J$ for the \dawave\ symmetry is indicated with a dotted line.}
\end{figure}

\begin{figure}[htb]
\includegraphics[scale = 0.65]{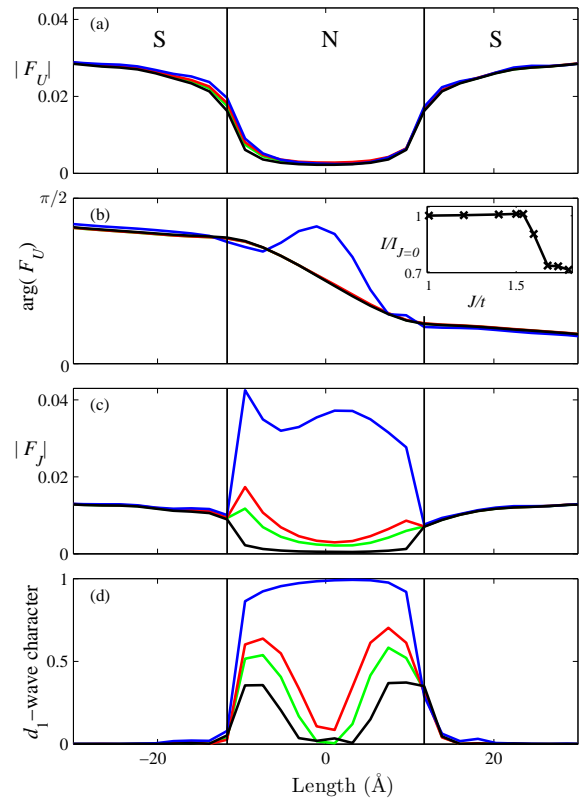}
\caption{\label{fig:b} (Color online) Zigzag SNS junction with $L = 10$ unit cells for four different values of $J$ in N: $J = 0$ (black), 3.5 (green), 3.75 (red), 4 eV (blue) and with $\tilde{\mu}(N) = 0.7$~eV. Absolute value (a) and phase (b) of the on-site pair amplitude $F_U$, absolute value of the SB pair amplitude $\Delta_J$ (c), and $d_1$-wave character of $F_J$ (d). Remaining character is $s$-wave. Vertical lines indicated the interfaces. Inset in (b) shows how the current changes with $J$ relative to the current at $J = 0$.}
\end{figure}

Figure \ref{fig:azz} shows typical results for a doped zigzag junction for different lengths $L$. Here $\Delta_J$, along with its character, in the N region in a SNS junction is shown together with $\Delta_J$ for a slab of the same length. The slab has been left with dangling bonds at the surfaces. As can be seen, despite the presence of the $\Delta_U$ superconducting state in the contacts, $\Delta_J$ is not enhanced in an SNS junction over the value in the slab. In addition, $\Delta_J$ is always suppressed in finite-size zigzag slabs compared to the bulk value. In terms of symmetries, we see that even the smallest slabs have a \dawave\ symmetry. 
%
%
%
While we would expect a $d_1+id_2$-wave symmetry in the bulk, the preference for the \dawave\ for finite-size zigzag slabs can be understood when considering that bonds 2 and 3 are equivalent whereas bond 1 is not for this interface. Since $d_1 \propto (2,-1,-1)$ it becomes energetically favored over the $d_2$-wave solution where bonds 2 and 3 instead have a $\pi$-phase difference.
In fact, this surface/interface effect last for the longest zigzag slabs ($\sim 500$~\AA) we can model. The only noticeable effect of the superconducting contacts in a zigzag junction is the \swave\ symmetry of $\Delta_J$ for weak SB pairing. However, as seen in the inset of Fig.~\ref{fig:b}(b), the \swave\ state is always too weak to influence the Josephson current through the junction. This weak extended \swave\ state appears only when $J$ and \chempot\ are too weak to cause a notable ($d$-wave) superconducting state in N and in fact only for finite doping levels. In undoped or barely doped graphene, $\Delta_J$ will always have $d$-wave character for any $J$ (including $J = 0$), but of course, the amplitude of the order parameter is (vanishingly) small. 
So despite the possible coupling between the S region on-site \swave\ and the extended \swave\ state of the SB pairing, the lowest energy state when SB pairing develops in the junction is still a $d$-wave state, as in the bulk. This applies even for very small junctions.

\begin{figure}[htb]
\includegraphics[scale = 0.65]{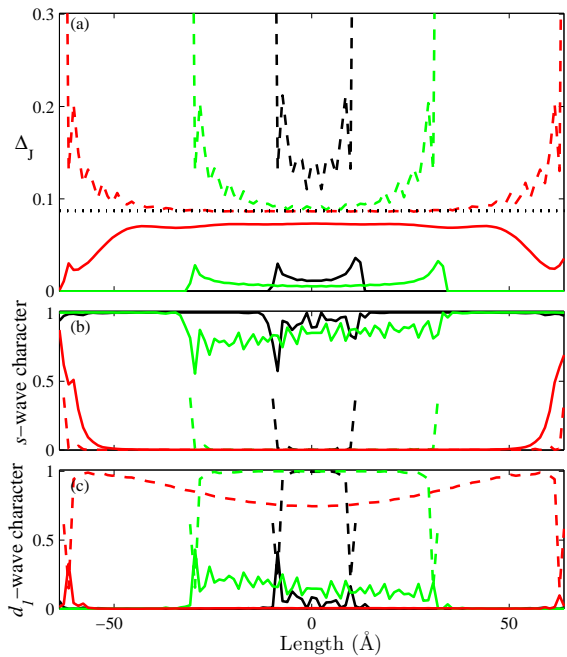}
\caption{\label{fig:aac} (Color online) (a) $\Delta_J$, (b) $s$-wave, and (c) $d_1$-wave character of $\Delta_J$ for armchair slabs (dashed) and SNS junctions (solid) for $L = 18$ (black), 52 (green), and 104 (red) unit cells when $J = 4$~eV and $\chempot({\rm N}) = 0.7$~eV. The bulk value of $\Delta_J$ for the $d_1+id_2$-wave symmetry is indicated with a dotted line.}
\end{figure}

For the armchair interface Fig.~\ref{fig:aac} shows the corresponding picture. There are a few differences compared to the zigzag interface.
%
First, while even the longest slab has $d_1$-wave symmetry at the surfaces, it goes toward $d_1+id_2$-wave symmetry ($d_1+id_2 = 50\%~d_1$-wave character and 50\% $d_2$-wave character) in the interior. This can be understood when examining the armchair interface more closely. Here, bond 2 in cell ${\bf i}$ is only equivalent to bond 3 in cell $({\bf i+1})$ and only if the paring potential parameters $U$ and $J$ are not changed in between the two cells. This is a much weaker symmetry condition than for the zigzag interface. 
Second, $\Delta_J$ is enhanced in finite-sized slabs compared to the bulk. This is because the value of $\Delta_{J{\bf a}_1}$ is very high at the surface. We have checked that this is not due to a localized surface state on this bond. If one was to suppress this value at the surface, $\Delta_J$ for a finite-sized slab will again resemble the value in a SNS junction. Since the value of $\Delta_{J{\bf a}_1}$ is not allowed to be as high at the interface with the S region as it is at a dangling bond surface the size of $\Delta_J$ in the SNS junction can again be understood in terms of the finite-size suppression in a simple slab. The symmetry choice for the slab is independent of any suppression of $\Delta_{J{\bf a}_1}$ at the surface.
Third, the \swave\ state lasts for somewhat longer junctions, or, alternatively larger $J$-values, in an armchair SNS junction, though its effect is similarly negligible on the Josephson coupling due to its smallness.
Fourth, for intermediate lengths $L$ the armchair SNS junction shows \dbwave\ symmetry while for even longer junctions (not shown in figure) there is a $d_1+id_2$-state in all of N. 
Thus the only difference between an armchair slab and a SNS junction in terms of $\Delta_J$ is the extended \swave\ in short junctions and the choice of $d_2$- or $d_1+id_2$-wave symmetries when the slab instead shows \dawave\ in the whole junction or only at the surface, respectively. 

Figures \ref{fig:azz} and \ref{fig:aac} establish the close connection between a simple slab and the N region in a SNS junction  for the SB order parameter but we are also interested in the development of the on-site pairing amplitude and the supercurrent through a SNS junction with varying $J$.
Figures \ref{fig:b}(a) and \ref{fig:b}(b) show how these properties develop with increased $J$ in a zigzag SNS junction. 
%
We clearly see that for all $J$, causing either extended \swave\ or \dawave\ symmetries in N, there is no effect on the amplitude of the on-site pairing amplitude $F_U$; its proximity effect depletion in S and leakage into N is unaffected by any SB correlations or superconducting state in N. Interestingly, the proximity length of the SB pairing as displayed in Fig.~\ref{fig:b}(c) is quite different from that of the on-site pairing. The SB pairing amplitude rises very abruptly at the interface and no effect of $J$ is seen in the S regions. As seen in Fig.~\ref{fig:azz}, the proximity length for the SB pairing is quite short at a surface as well, but the difference is even more pronounced in the SNS structure.
When $J$ is strong enough to cause a \dawave\ state in N, we see a clear signature on the phase of $F_U$. The total phase drop over the junction remains the same, and thus the $I$ vs $\Delta \phi$ relationship is unchanged, but there is a significant `bulge'\ developing in N for arg($F_U$). 
The inset in Fig.~\ref{fig:b}(b) shows the development of the Josephson current with increased $J$ compared to when $J=0$. For all \swave\ states there is a small increase in the current (less than 1\%) caused by the coupling between the extended \swave\ in N and $\Delta_U$ in S. But since $\Delta_J$ is very small in this situation the increase in current is also quite small.
Once the \dawave\ state develops we see a significant {\it decrease} in the current. This can be explained by the following argument: the \dawave\ state has according to Fig.~\ref{fig:scop} zero overlap with a \swave\ state in a SS$^\prime$ junction, so there is no coupling between the two condensates, at least not to first order. Excluding possible contributions from higher order terms in this coupling, the current carried through the junction can only have two origins; quasiparticle tunneling and Andreev tunneling in which an incoming Cooper pair on the interface in S is transformed into an outgoing electron and an incoming hole on the interface in N. These are the same physical processes present in a regular SNS junction and the latter process should dominate at the low temperatures we are considering. Both of these tunneling processes are sensitive to changes in the normal or, in the case of a superconducting state, quasiparticle DOS in N. For $J = 0$ the N region is just doped graphene which has a metallic DOS. When the \dawave\ state develops, the DOS is changed to  a nodal spectrum since the \dawave\ has nodes on the Fermi surface. This effectively decreases the available DOS for the above mentioned tunneling processes and the current will accordingly decrease. The same mechanism will decrease the tunneling current in the case of a \swave\ state in N as well, but in this case the superconducting energy gap is very small and the decrease is also counteracted by the Josephson coupling between the two different \swave\ order parameters. The final result is a marginal increase in the current instead.

The corresponding picture for an armchair junction is very similar, except for the symmetry of the $d$-wave state. As commented above, for intermediate length junctions and strong enough $J$, a \dbwave\ state will develop in the armchair junction. This state has zero overlap with the S region \swave\ state and we again see a drop in the current as for the zigzag junction. 
For the longest junctions we could study the zigzag junction is still in a \dawave\ state whereas the armchair has a $d_1+id_2$-wave symmetry. This state has a finite overlap of the $d_1$-wave part with the S region \swave\ state but counteracting this increase in the current is the fact that the $d_1+id_2$-wave state has a full gap on the Fermi surface and, for the $J$-values we are considering, also a considerable size of the gap. This effectively diminishes any quasiparticle and Andreev contribution to the current. The final result is in fact a reasonably large decrease in the total current compared to when $J = 0$. 
For the armchair interface the `bulge'\ structure on the phase of $F_U$ is present for both the \dbwave\ and the $d_1+id_2$-wave order parameters. 

The choice of symmetry for the SB order parameter in N is interesting. For zigzag junctions, clearly the surface preference of bond 2 = bond 3 $\neq$ bond 1 will dictate a \dawave\ state for even quite long junctions. On the other hand for armchair junctions, a slab will always have a \dawave\ symmetry at the surface but in a SNS junction the symmetry is either $d_2$- or $d_1+id_2$-wave in the whole N region.  It is almost as if the system has the lowest free energy when it can, to as large extent as possible, avoid overlap between the S region \swave\ and the N region $d$-wave order parameters.

From the results in Figs.~\ref{fig:azz}-\ref{fig:b} we can draw the following three conclusions of the effect of SB correlations in clean graphene SNS Josephson junctions: (1) SB pairing in a SNS junction is going to suffer from finite-size suppression, in which a larger $J$ (or \chempot) than in the bulk is necessary in order to achieve any SB pairing. This clearly answers our first question as to whether a conventional SNS junction can enhance the effect of SB correlations or not. (2) Even with a large enough $J$ such that a SB superconducting state develops in N, the junction is going to resemble the situation `bulk-meets-bulk'\ in terms of the lack of induced effects in between the different order parameters. (3) The current through the junction will depend on both the possible overlap of the different order parameters in S and N but also to a significant degree on the DOS of quasiparticles in N. 

Since these conclusions, with the exception of the possible symmetries for the $d$-wave state in N, are identical for the zigzag and the armchair interfaces we expect that they will be true for any clean chiral interface SNS junction. The issue of choice for the $d$-wave state will require a detailed knowledge of the interface structure.
%
%
\subsection{SNS junctions with rough interfaces}
\begin{figure}[htb]
\includegraphics[scale = 0.65]{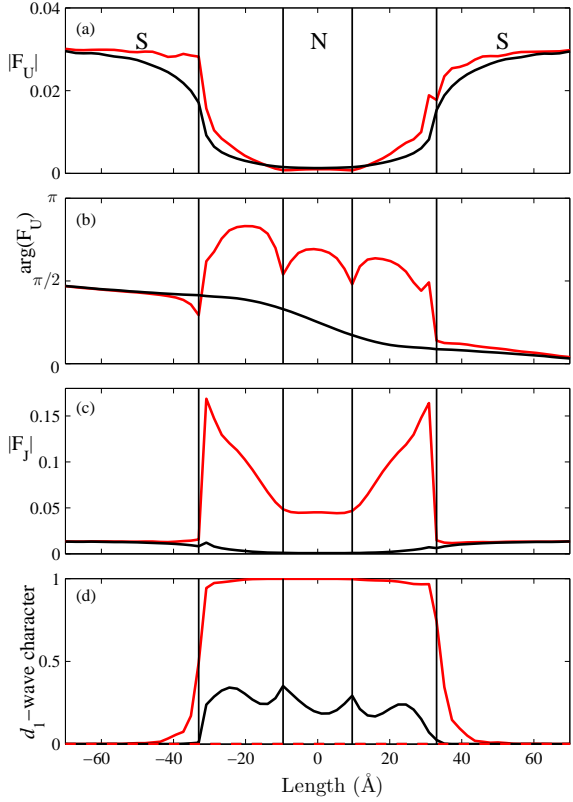}
\caption{\label{fig:dlin} (Color online) Zigzag SNS junction with linear drop in doping profile over 10 unit cells from $\chempot({\rm S}) = 1.5$~eV to $\chempot({\rm N}) = 0.7$~eV for J = 0 (black), 4 eV (red). Absolute value (a) and phase (b) of the on-site pair amplitude $F_U$, absolute value of the SB pair amplitude $F_J$ (c), and $d_1$-wave character of $\Delta_J$ (d). Remaining character is $s$-wave. Vertical lines indicate the SN interfaces as well as where the linear doping profile drop ends.}
\end{figure}

After studying the clean interface SNS Josephson junctions we are still left with the question if it is possible to induce a coupling between the on-site \swave\ $\Delta_U$ and the SB $d$-wave $\Delta_J$ by making the interface less idealistic. We first investigate the effect of the sharp doping level drop at the interface. It is reasonable to assume that some doping is going to leak into N from the higher doped S region causing either the change \chempot(S) $\leftrightarrow$ \chempot(N) to take place at another site than where $U \neq 0 \leftrightarrow J \neq 0$ or, even more realistically, assume a steady, nonabrupt, drop from \chempot(S) to \chempot(N) over a few unit cells. We have investigated both possibilities and they show similar features. Figure  \ref{fig:dlin} shows the situation where the chemical potential is allowed to drop linearly from the value in S to the value in N over 10 unit cells.
%
As seen, a finite $\Delta_J$ gives rise to some induced on-site pairing amplitude $F_U$. Note however that the much larger value of the SB pairing amplitude $F_J$ in the region where the doping drops take place is at least primarily due to the effectively higher chemical potential in that region and not because of coupling to $F_U$.
A similar effect can be seen even with a sharp drop in chemical potential as long as it does not coincide with the change from on-site to SB pairing potential. This behavior is different from the clean interface and indicates that the big FLM at the interface is a key component in the suppression of coupling between the two order parameters. But, even with the removal of this constraint, there is still a finite-size effect present, i.e.~a larger $J$ than in the bulk is still needed in order to induce a finite $\Delta_J$ and hence to see an increased $F_U$. Also, the symmetry choice for $\Delta_J$ does not change and the intriguing `bulge'\ structure is still present. In terms of the Josephson current through the junction, it is obviously increased when $F_U$ is increased, in this particular case by about 10\% compared to when $J = 0$.

\begin{figure}[htb]
\includegraphics[scale = 0.65]{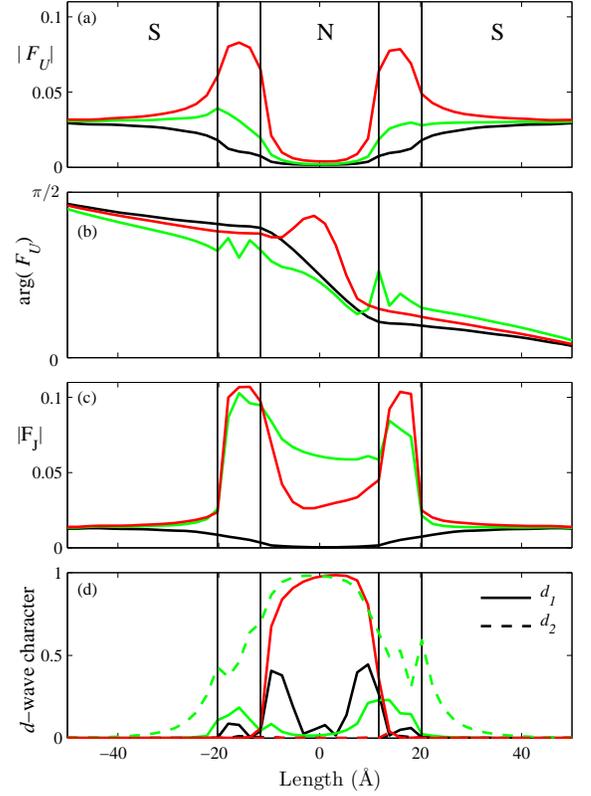}
\caption{\label{fig:drough} (Color online) Rough interfaces for zigzag SNS junction for J = 0 (black), J = 4 eV (red, green) with a sharp doping profile drop to $\chempot({\rm N}) = 0.7$~eV at the innermost vertical lines. In the interface regions (between vertical lines at each interface) $U \neq 0$ for all cells and $J \neq 0$ for 50\% of the cells (red) and $U = 0, J \neq 0$ or $U \neq 0, J = 0$ in equal mixture (green).
Absolute value (a) and phase (b) of the on-site pair amplitude $F_U$, absolute value of the SB pair amplitude $F_J$ (c), and $d$-wave character, $d_1$ (solid), $d_2$-wave (dashed) of $\Delta_J$ (d). Remaining character is $s$-wave.}
\end{figure}

Figure \ref{fig:dlin} shows that it is possible to get induced pairing amplitude effects between the on-site \swave\ state and a SB $d$-wave state. An additional question is if it, by interface engineering, is possible to choose the $d$-wave state in N that has a nonzero overlap with the \swave\ state at the interface, and thereby allow for coupling between the two supercurrents. For the zigzag interface this would mean developing a \dbwave\ symmetry in N. Figure \ref{fig:drough} shows that this is indeed possible when care is taken in the interface region. More specifically, the strong symmetry between bond 2 and 3 has to be broken. We achieve this by doubling the unit cell in the direction along the interfaces and setting $U$ and $J$ to different values on the two $f$-atoms in this new extended cell.
%
Here the green (light grey) curve represents a junction where the \dbwave\ symmetry is chosen and the current is also increased by $50\%$. There is also a reasonable inducement of on-site pairing amplitude from the SB pairing. In order to establish that the current increase is not just due to this increase in on-site pairing but also due to coupling of the two condensates, we compare the results with a system where $J \neq 0$ is allowed to overlap with $U \neq 0$. In this system, plotted in red (grey), there is a very strong coupling between the on-site \swave\ and a SB \swave\  in the interface region. Despite the much larger induced $F_U$ in this latter case the current increase is only around 20\%. This proves that the majority of the current increase in the \dbwave\ case comes from the overlap of order parameters and the flow of supercurrent between them.

Interestingly we also see that the `bulge'\ on arg($F_U$) disappears for the zigzag junction when the \dbwave\ symmetry is chosen. We therefore speculate that the origin of the `bulge'\  is the response of the on-site order parameter to the lack (or in the case of $d_1+id_2$-wave, partial lack) of order parameter overlap and/or the suppression in supercurrent that effectively follows from this.

\subsection{LDOS and the existence of ZES}
\begin{figure}[htb]
\includegraphics[scale = 0.65]{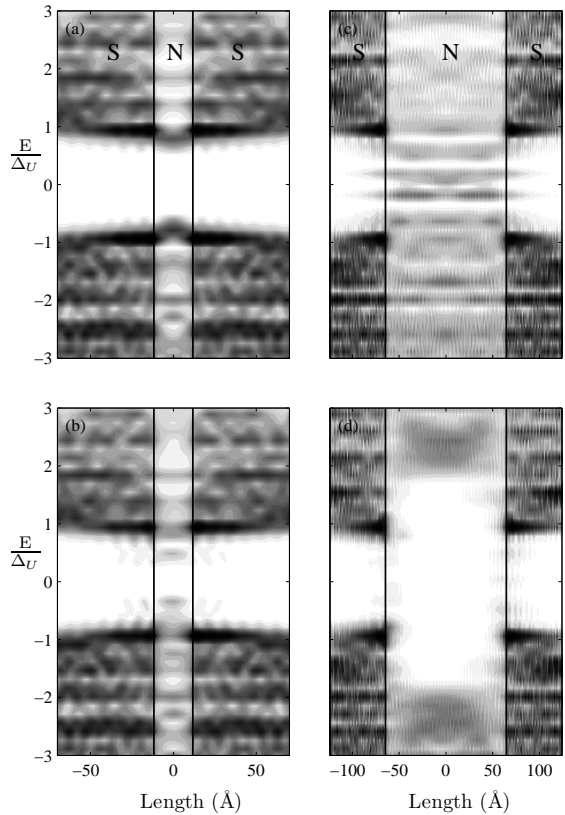}
\caption{\label{fig:eldos} LDOS for zigzag junction with $L = 10$ unit cells and $J = 0$ (a), $J = 4$ ~eV (b) and for armchair junction with $L = 104$ unit cells and $J = 0$ (c), $J = 4$ ~eV (d), all with $\chempot({\rm N}) = 0.7$~eV. Zigzag junction (b) has $d_1$-wave symmetry in N which has nodes in the gap function, armchair junction (d) has $d_1 + id_2$-wave symmetry which is fully gapped. Black color corresponds to 2~states/eV/unitcell. Vertical lines indicates the interfaces.}
\end{figure}

Finally we discuss the LDOS for different junctions. Figure \ref{fig:eldos} shows the LDOS for a clean zigzag SNS junction with $J=0$ (a) and $J \neq 0$ with a \dawave\ symmetry in N (b) and for a clean armchair SNS junction with $J=0$ (c) and $J \neq 0$ with a $d_1+id_2$-wave symmetry in N (d). 
%
First it is clear that the abrupt drop in the effective chemical potential parameter at the SN interface gives an abrupt drop even in the electron filling of the bands. This is seen in the lighter color in the N region which corresponds to a lower DOS and is the result of the N region being closer to the zero DOS Dirac point. 
Second we see that for a relatively short junction (a) the energy gap is not depleted inside N when $J = 0$ whereas for a longer junction (c) it is to a large extend fully depleted due to the length of the junction. When $J \neq 0$ we see in (b) the appearance of the nodal quasiparticle spectrum characteristic of the \dawave\ superconducting state whereas in (d) we see a large energy develop due to the fully gapped $d_1+id_2$-wave state. 

There is some tendency for interfacial states in the gap in Fig.~\ref{fig:eldos}(d) but the DOS is still very low and no localized states are actually formed.
This however raises the question of the existence of zero energy states (ZESs) at surfaces or interfaces. For high-$T_c$ superconductors it was shown relatively early on that midgap surface states will exist at certain surfaces.\cite{Hu94} Later on this was extended to also include the formation of ZESs at various interfaces between superconductors with unconventional order parameters (see Ref.~[\onlinecite{Kashiwaya00}] and references therein). 
The surface state can be viewed as a bound state formed in a N layer on top of the superconductor in the limit where the thickness of N goes to zero. Bound states are formed in this layer due to the retro-reflectivity of the Andreev reflection which causes an electron traveling in the N layer to form a closed trajectory. For conventional superconductors these states coincides with the de Gennes-Saint-James bound states.\cite{deGennes63} However, for unconventional order parameters the two Andreev reflections involved in this process will take place via different order parameters. In order for a surface state to form at zero energy it is only necessary that $\Delta(\theta)$ and $\Delta(\pi-\theta)$, where $\theta$ is the angle of incidence for an electron quasiparticle at the surface, have different signs.\cite{Kashiwaya00} 
In our case the necessary sign change in the order parameter is true for the \dbwave\ symmetry at a zigzag surface and the \dawave\ symmetry at an armchair surface where the latter case appears naturally for a clean interface, as seen in Fig.~\ref{fig:aac}. Despite this we were not able to detect any trace of a ZES at the armchair surface. We attribute the absence of a ZES to the local symmetry character changes in the surface. Within the first few unit cells of the armchair surface the character of the order parameter has both strong $s$- and \dbwave\ components, which neither give rise to a ZES. In fact, the only time we were able to see the signature of a ZES was for a zigzag surface when we forced the symmetry to be \dbwave\ in the whole slab. A similar enforcement of \dawave\ symmetry in a whole armchair slab leads to a depleted pair potential at the surface, and consequently also a lack of a ZES. It might be worth emphasizing that we are here dealing with rather heavily doped graphene samples so we do not expect the rather peculiar specular Andreev reflection possible near the Dirac point\cite{Beenakker06} to possibly play a role here.

ZESs in a SS$^\prime$ junction have a similar origin to the surface states. In fact, in the tunneling limit the existence of a ZES is determined independently at each of the two superconductor surfaces. In the opposite, fully transparent limit, a sign change in the order parameters has to be present across the junction instead. In the region of intermediate transparency a linear combination of these two conditions applies.\cite{Kashiwaya00}
For a square lattice with $d_{x^2-y^2}$-wave superconductor junctions treated within the TB-BdG formalism, ZESs were reported for several interfaces including the \{110\} interface.\cite{Shirai03} However, it was also found that when using a lattice model the appearance of a ZES is sensitive to Friedel oscillations in the wave function. These can cause destructive interference between different surface lattice sites, leading to the disappearance of the ZES.
For our graphene junctions with conventional \swave\ on one side and $d$-wave on the other side, the same criterion as for a surface for seeing a ZES applies, independent of the transparency of the interface. For the junctions at hand, only the rough interface \dbwave\ symmetry zigzag junction would qualify but, as in the surface case, the character of the order parameter is very mixed at the interface, and we do not see any sign of a ZES. 

Finally, let us comment on the case of the time-reversal symmetry breaking $d_1+id_2$-wave state. This state will not have any ZES, but it can still have subgap localized states, which can be viewed as a splitting of the zero energy peak.\cite{Kashiwaya00} However, apart from the small DOS at the interface in Fig.~\ref{fig:eldos}(d) we have not been able to see any signs of subgap states. The Andreev reflection process described above to give rise to the ZES at the surface of the superconductor will in fact in this case produce a net current along the interface. This is, however, beyond the scope of this paper.
%
\section{Conclusions}
We have investigated the possibility of enhancing the intrinsic nearest neighbor spin-singlet (SB) correlations, present in all  $p\pi$-bonded planar organic molecules, in graphene by constructing a graphene SNS Josephson junction with conventional \swave\ superconducting contacts. For strong enough SB coupling constant $J$ (and doping level) a $d_1+id_2$-wave superconducting state develops in the bulk. However, we have found that in a finite-sized junction, as in a finite-sized slab, a larger $J$ than in the bulk is needed  to achieve superconductivity. In addition, the SB superconducting state will still have $d$-wave symmetry despite the nonzero coupling between the SB extended \swave\ state to the extrinsic \swave\ state induced from the superconducting contacts. 
Using a realistic estimation of  $J$, it would be necessary to have a very high doping level in the N region in order to develop a $d$-wave state. While such high doping levels cannot as of present be achieved by traditional gating of the graphene, there might be a possibility of using chemical doping, such as is assumed for the S regions in this work. However, care has to be taken to not change any other physical characteristics of the graphene other than the Fermi level. 
An interesting side remark is that, for a dangling bond short armchair slab, it might be possible to see a SB superconducting state for lower doping levels than in the bulk thanks to a strong singlet formation on the surface bond.

We have also found that interface effects are very important for the choice of $d$-wave symmetry. For even the longest clean zigzag junctions we could model a \dawave\ symmetry is favored whereas for the clean armchair junction \dbwave\ symmetry is favored in shorter junctions but the $d_1+id_2$-wave state is reached for sufficiently long junctions.
Also for the clean junctions, the on-site \swave\ pair amplitude in S and the $d$-wave in N do not influence each other and the situation can to good approximation be described as a `bulk-meets-bulk'\ junction. For the \dawave\ in a zigzag junction or the \dbwave\ in an armchair junction there is also no coupling or overlap between the $d$-wave in N and the \swave\  in S, inhibiting any Josephson coupling in the junction. In fact, the supercurrent {\it decreases} when the $d$-wave develops in N as quasiparticle tunneling and Andreev reflection are suppressed due to the nodal spectrum developing in the quasiparticle DOS in N for the $d_1$- and $d_2$-wave. The $d_1+id_2$-wave has a finite overlap of one of its components with the S region \swave\ state but the quasiparticle DOS in N is fully gapped and, when both effects are combined, we have found that the net effect is still a decrease in the current.
We have also reported on the development of a `bulge'\ in the phase of the on-site pair amplitude when a $d$-wave develops in N. This structure is pronounced but will not influence the current vs phase relationship in the junction and it is only present when (part of) the $d$-wave state lacks coupling with the on-site \swave. 
For the clean zigzag or armchair interfaces this `bulge'\ feature is, except for a  $ <10\%$ decrease in current, the only sign of a $d$-wave in N in the usual Josephson junction characteristics. The structure is however pronounced and might be possible to detect with a phase sensitive measurement inside N.

In order to facilitate coupling between the two pair amplitudes we have also investigated interfaces with smoother doping profiles and atomic scale interface roughness and been able to see an increased on-site pair amplitude, and thus higher supercurrent, when the $d$-wave state develops in N. However, in order to drastically increase the current, the opposite $d$-wave symmetry than the one found naturally in clean junctions has to be favored. We have shown that this is possible when including interface roughness and the current has been shown to increase with $50\%$ in one such case. This would be a clear signal for a developing superconducting state due to SB correlations.

Because of computational constraints we have only investigated the two simplest interfaces, zigzag and armchair, but in a general experimental junction it is likely that a chiral and also nonsmooth interface is present. Due to the near degeneracy in energy of the two different $d$-wave solutions and their complex combinations, it is impossible to say, without detailed knowledge about the interface, which symmetry will be favored in a general junction and as seen here, the choice of symmetry will have a big influence on the properties of the junction. However, as an experimental junction will at least experience some doping leakage into N and not be perfectly clean on the atomic level, it will, with high enough doping in the N region to induce a SB superconducting state, show an increased Josephson current due to an increased on-site pairing potential, and possibly even a significant increase due to a particular choice of $d$-wave symmetry in N.

Despite the unconventional order parameters and the theoretical existence of zero energy states (ZESs) at a surface or interface of such superconductors, most junctions and slabs we have investigated do not show any signs of ZESs in the gap. We attribute this to the delicate symmetry character mixing at surfaces and interfaces. Therefore, it is unlikely that one can detect the presence of a $d$-wave state in an experimental system by searching for a zero energy peak in the LDOS or the zero bias conductance peak it is known to create.

Finally, let us point out the possible benefit of a SNS junction with $d$-wave contacts, made by depositing, for example, a high-$T_c$ material on graphene. As has been clear from this study, the biggest problem with conventional contacts is the zero coupling between the \swave\ and the $d$-wave states in the bulk. For rough interfaces, coupling can to some extent be obtained but the fundamental symmetry difference is still a big obstacle. While it remains to be studied, we believe that $d$-wave contacts could potentially greatly enhance the effect of the intrinsic SB correlations. A corresponding study has been made on the square lattice with $d$-wave contacts and a N region with $d$-wave correlations but $T_c < T$ and pronounced effects in proximity effect and current were seen.\cite{Covaci06} For graphene there would also be the additional issue of having multiple $d$-wave symmetries which complicate the picture, but if any effects could be seen, these would experimentally prove the existence of SB correlations in graphene.
%
%
%
\begin{acknowledgments}
A.M.B.-S. acknowledges partial support from DOE under Contract No.~DE-AC02-76SF00515.
\end{acknowledgments}


\end{document}